\documentclass[12pt]{iopart}

\usepackage{graphicx}
\usepackage{bm}
\usepackage{iopams}

\begin{document}

\title{Hydrodynamic coupling between two fluid membranes}

\author{Sanoop Ramachandran and
Shigeyuki Komura
}

\address{
Department of Chemistry,
Graduate School of Science and Engineering, \\
Tokyo Metropolitan University, 
Tokyo 192-0397, Japan
}
\ead{komura@tmu.ac.jp}

\begin{abstract}
The coupled in-plane diffusion dynamics between point-particles 
embedded in stacked fluid membranes are investigated.
We calculate the contributions to the coupling longitudinal and 
transverse diffusion coefficients due to particle motion within 
the different as well as the same membranes.
The stacked geometry leads to a hydrodynamic coupling between the 
two membranes.
\end{abstract}


Biological membranes are fundamental to the existence of life
with their ability to separate the ``in'' and ``out'' of a
cell.
Typical membranes are composed of lipid molecules
which can spontaneously self-assemble into a fluid bilayer
structure~\cite{alberts}.
Under normal physiological conditions, finite temperature
induces Brownian motion of membrane constituents, 
resulting in diffusive transport.
In the seminal paper by Saffman and Delbr\"uck~\cite{saffman-75}, 
the diffusion coefficient of a rigid protein in a membrane
was calculated. 
Following this work, there have been theoretical studies on 
diffusion of a rigid disk on supported membranes~\cite{evans-88},
a liquid domain in monolayers~\cite{lubensky-96,sanoop-drag-10},
and a rod on immersed membranes~\cite{levine-04,levine-04b}
or on Langmuir monolayers~\cite{fischer-04}.

The present work focuses on the coupling of diffusion
dynamics between two fluid membranes through an intervening 
bulk fluid.
It was reported that model experimental systems in which two 
lipid membranes are stacked on a substrate exhibit correlated 
dynamics~\cite{kaizuka-04}.
This planar geometry has become favorable to study the membrane 
dynamics which are otherwise not possible in vesicles~\cite{groves-07}.
The coupling effect between two membranes can be important 
in biological systems with large concentration of cells such as 
in tissues.
Other examples are Gram-negative bacteria which enclose a 
periplasmic space with an approximate width of about 15 to 20 nm 
between their inner and outer lipid bilayers~\cite{alberts}.
Highly folded membranous organelles such as Golgi apparatus
also correspond to a situation in which membranes come in close 
proximity to each other.
In all these cases, it is very relevant to consider the hydrodynamic 
coupling between two biomembranes.

Hydrodynamic models of membranes exploit their fluid nature, 
in which the membranes are assumed to be two-dimensional (2D)
viscous fluid sheets embedded in a three-dimensional (3D) solvent.
One such investigation involved the calculation of the correlated 
diffusion of proteins embedded in a membrane immersed in an 
unbounded fluid~\cite{oppenheimer-09} or a membrane adjacent to 
a solid support~\cite{oppenheimer-10}.
In this Communication, we report on the coupling diffusion 
coefficients of particles embedded in such a stacked membrane system.

\begin{figure}
\begin{center}
\includegraphics[scale=0.35]{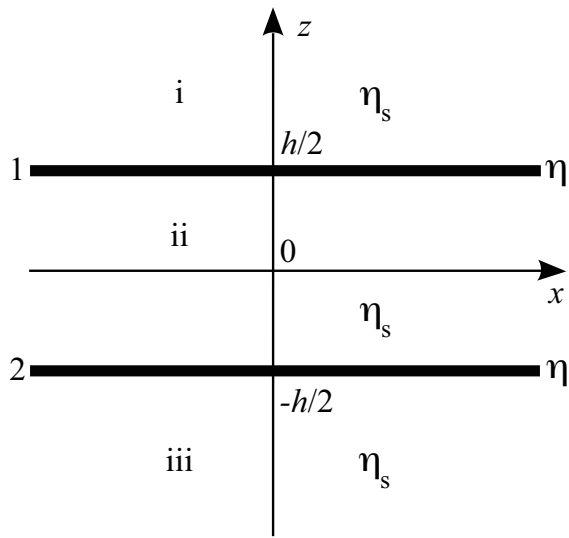}
\caption{Schematic picture showing a set of stacked fluid membranes
(solid thick lines) 
having 2D viscosities $\eta$ located at $z=\pm h/2$.
The two membranes are denoted by the labels 1 and 2.
The solvent of 3D viscosities $\eta_{\rm s}$ are labeled by
the regions i, ii and iii.
}
\label{fig:doublememb}
\end{center}
\end{figure}

We first establish the governing equations. 
It is assumed here that the membranes are infinite planar sheets of fluid.
The persistence length is defined as the length over which the correlations 
of the normal vector decay to zero~\cite{degennes-taupin-82}.
For a lipid bilayer membrane with bending rigidity of about 10 
$k_{\rm B}T$, the persistence length turns out to be much larger 
than the size of a typical cell.
Hence we neglect out-of-plane fluctuations of the membrane for 
mathematical simplicity.
The fluid membranes are embedded in a bulk solvent such as water or 
suitable buffer solution.
As shown in figure~\ref{fig:doublememb}, the membranes are fixed in the 
$xy$-plane at $z=\pm h/2$.
Let ${\bf v}^{(i)}({\bf r})$ be the 2D velocity of the membrane fluids.
Here the index $i = 1,2$ represents the two membranes,
and the 2D vector ${\bf r}=(x,y)$ represents a point in the planes 
of the membranes.
We work in the low-Reynolds number regime of the membrane hydrodynamics
so that the inertial effects can be neglected.
This allows us to use the 2D Stokes equations 
\begin{equation}
\eta \nabla^2 {\bf v}^{(i)} - \nabla p^{(i)}
+ {\bf f}_{\rm s}^{(i)} + {\bf F}^{(i)} =0,
\label{eqn:2Dstokes}
\end{equation}
along with the incompressibility condition,
\begin{equation}
\nabla \cdot {\bf v}^{(i)}=0.
\label{eqn:2Dcompress}
\end{equation}
Here $\nabla$ is a 2D differential operator,
$\eta$ is the 2D membrane viscosity (same for both the membranes),
$p^{(i)}({\bf r})$ the 2D in-plane pressure,
${\bf f}^{(i)}_{\rm s}({\bf r})$ the force exerted on the membrane 
by the surrounding fluid,
and ${\bf F}^{(i)}({\bf r})$ is any other force acting on the membrane.

The solvent regions are denoted by the index $j=$ i, ii, iii. 
The velocities and pressures in these regions are written as 
${\bf v}^{(j)}({\bf r},z)$ and $p^{(j)}({\bf r},z)$, respectively.
We assume that the solvent in the three regions have the same
3D viscosity denoted by $\eta_{\rm s}$. 
The solvent inertia is neglected and hence it also obeys 
the 3D Stokes equations 
\begin{equation}
\eta_{\rm s} \tilde{\nabla}^2 {\bf v}^{(j)} - 
\tilde{\nabla} p^{(j)} = 0,
\label{eqn:3Dstokes}
\end{equation}
where $\tilde \nabla$ represents a 3D differential operator.
Similar to the fluid membrane, the solvent in all the regions
are considered to be incompressible 
\begin{equation}
\tilde{\nabla} \cdot {\bf v}^{(j)}  = 0.
\label{eqn:3Dcompress}
\end{equation}
The presence of the surrounding solvent is important because it 
exerts force on the liquid membranes.
The force on membrane 1, indicated as ${\bf f}^{(1)}_{\rm s}$ 
in (\ref{eqn:2Dstokes}), is given by the projection of 
$({\bm \sigma}^{({\rm i})} - {\bm \sigma}^{({\rm ii})})_{z=h/2} 
\cdot \hat{{\bf e}}_z$
onto the $xy$-plane of the membrane.
Here $\hat{{\bf e}}_z$ is the unit vector along the $z$-axis, 
and ${\bm \sigma}^{(j)}$ are the stress tensors 
\begin{equation}
{\bm \sigma}^{(j)} = 
- p^{(j)} {\bf I} 
+ \eta_{\rm s} \left[ \tilde{\nabla} {\bf v}^{(j)}
+ (\tilde{\nabla} {\bf v}^{(j)})^{\rm T} \right].
\label{eqn:stress}
\end{equation}
In the above, ${\bf I}$ is the identity tensor and the superscript
``T'' indicates the transpose.
Similarly, the force on membrane 2, is given by the projection of 
$({\bm \sigma}^{({\rm ii})} - {\bm \sigma}^{({\rm iii})})_{z=-h/2} 
\cdot \hat{{\bf e}}_z$ on the $xy$-plane.
The general procedure is to first resolve (\ref{eqn:2Dstokes})
into components along ${\bf k}$ and perpendicular to it, where
${\bf k}$ is the 2D wave vector in Fourier space.
We then solve the resulting differential equations for the velocities.
Stick boundary conditions at the membrane-solvent interfaces
are imposed.
It is also assumed that the solvent velocities decay to zero at 
sufficiently large distances from the membranes.
A similar calculation for a single membrane confined between
two walls has been previously 
performed~\cite{sanoop-conc-10,sanoop-poly-10}.

Owing to the linearity of governing Stokes equations,
the in-plane velocity in membrane 1 can be obtained
in Fourier space as
\begin{equation}
v^{(1)}_\alpha[{\bf k}] = 
G^{(11)}_{\alpha\beta}[{\bf k}]F^{(1)}_\beta[{\bf k}] +
G^{(12)}_{\alpha\beta}[{\bf k}]F^{(2)}_\beta[{\bf k}].
\end{equation}
Here $G^{(11)}_{\alpha\beta}$ and $G^{(12)}_{\alpha\beta}$
($\alpha,\beta=x,y$) are the mobility tensors given by  
\begin{equation}
G^{\rm (11)}_{\alpha\beta}[{\bf k}] = 
\frac{1}{\eta \nu^2} \,
\frac{1+2 K +  \coth(KH)}
{K g(K,H)}
\left(
\delta_{\alpha\beta}- \frac{k_\alpha k_\beta}{k^2} 
\right), 
\end{equation}
\begin{equation}
G^{\rm (12)}_{\alpha\beta}[{\bf k}] = 
\frac{1}{\eta \nu^2} \,
\frac{ {\rm cosech}(KH)}
{K g(K,H)}
\left(
\delta_{\alpha\beta}- \frac{k_\alpha k_\beta}{k^2} 
\right),
\end{equation}
with 
\begin{equation}
g(K,H) = 1+ 2K(1+K)+(1+2K)\coth(KH),
\end{equation}
and $\nu \equiv 2 \eta_{\rm s}/\eta$. 
Notice that $\nu^{-1}$ represents the hydrodynamic screening length.
We have also used the definitions
$H = h\nu$ and $K = k/\nu$ with $k = |{\bf k}|$. 
By symmetry, a similar set of expressions can be written down 
for the membrane 2 also.
The above mobility tensors represent the hydrodynamic coupling 
between the two membranes.

Consider a pair of particles embedded in the membrane undergoing 
Brownian motion separated by ${\bf r}$.
For sufficiently large enough times, the particle displacements obey 
$\langle \Delta r_\alpha \Delta r'_\beta \rangle 
= 2 D_{\alpha\beta}({\bf r}) t$ where 
$\Delta r_\alpha$ represents the displacement of the first
particle and $\Delta r'_\beta$ represents that of the second 
particle.
In the above, the diffusion tensor for over-damped dynamics
is given by the Einstein relation 
$D_{\alpha\beta}=k_{\rm B}T G_{\alpha\beta}$ where 
$k_{\rm B}$ is the Boltzmann constant and $T$ the temperature.
We now apply these definitions to a double-membrane system.
The line of centers connecting any two particles in 
the membranes after projection on to the 2D plane 
can be taken to be along the $x$-axis without loss of generality.
Then we obtain the coupling \textit{longitudinal} diffusion 
coefficients as 
$D_{\rm L}^{(11)}({\bf r})= k_{\rm B} T G^{(11)}_{xx}(r\hat{e}_x)$
of two particles within the \textit{same} membrane,
and $D_{\rm L}^{(12)}({\bf r})= k_{\rm B} T G^{(12)}_{xx}(r\hat{e}_x)$
of two particles in \textit{different} membranes.
The coupling \textit{transverse} diffusion coefficients are
$D_{\rm T}^{(11)}({\bf r})=k_{\rm B} T G^{\rm (11)}_{yy}(r\hat{e}_x)$
and 
$D_{\rm T}^{(12)}({\bf r})=k_{\rm B} T G^{\rm (12)}_{yy}(r\hat{e}_x)$.
The longitudinal coupling diffusion coefficient is associated with
Brownian motion along the line of centers, while the transverse one 
is associated with motion perpendicular to the line of 
centers~\cite{oppenheimer-09,oppenheimer-10,sanoop-bulk-10}.
In the following, we discuss the above four diffusion coefficients 
sequentially.

\begin{figure}
\centering
\includegraphics[scale=0.45]{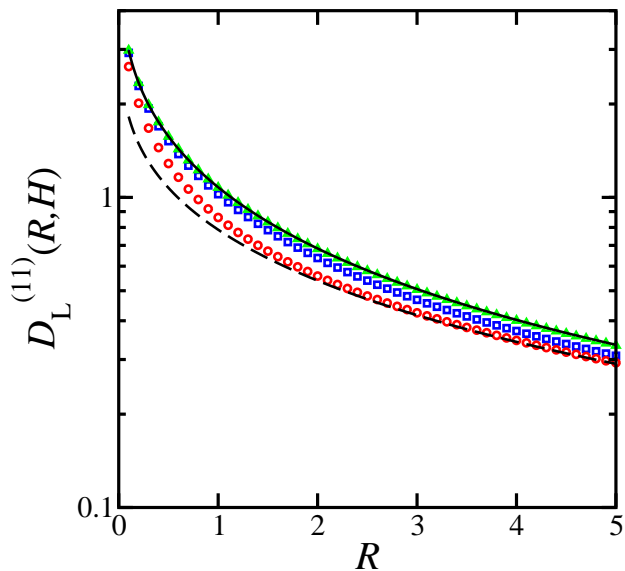}
\caption{
Dimensionless $D_{\rm L}^{(11)}(R,H)$ as a function 
of $R$ for various values of $H$. 
The red circles, blue squares and green triangles correspond 
to $H=0.1$, $1$ and $10$, respectively (color online).
The solid line corresponds to (\ref{eqn:DLSD}) representing 
the analytical expression for large $H$ limit.
The dashed line corresponds to the small $H$ limit.
}
\label{fig:DL1}
\end{figure}

The real-space expressions for the mobility tensors can be 
obtained by inverse Fourier transform.
Using the notation $R = r\nu$, we obtain the longitudinal 
coupling diffusion coefficient within the same membrane as
\begin{equation}
D_{\rm L}^{(11)}(R,H)
= 2\int_0^\infty {\rm d}K \,
\frac{1 + 2K +  \coth(KH)}{g(K,H)}\,
\frac{J_1(KR)}{KR},
\label{eqn:DL1}
\end{equation}
where $J_n(t)$ are the Bessel function of the first kind of order $n$.
Hereafter, all the diffusion coefficients are scaled by a factor 
$k_{\rm B}T/(4\pi\eta)$ in order to make them dimensionless.
In figure~\ref{fig:DL1}, we plot $D_{\rm L}^{(11)}(R,H)$ as a
function of $R$ for various $H$ following numerical integration.
It is a monotonically decreasing function of $R$. 
The red circles ($H=0.1$), blue squares ($H=1$) and green triangles ($H=10$)
represent the intermediate membrane separations respectively.
These symbols represent the same $H$ values in the rest of the figures
in this Communication.

\begin{figure}
\centering
\includegraphics[scale=0.45]{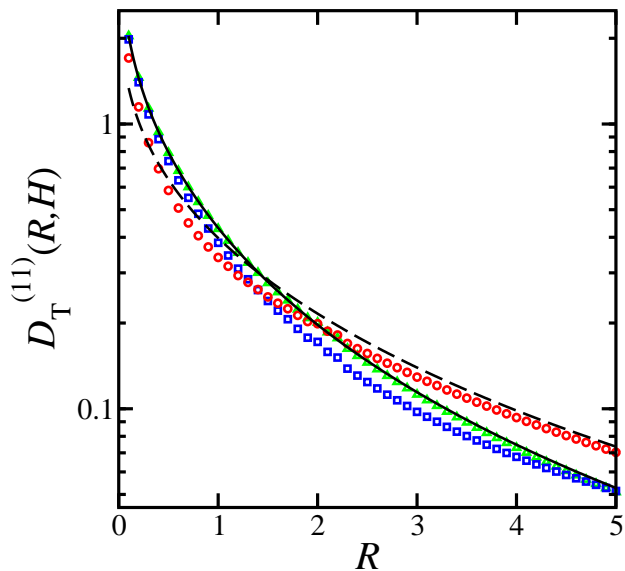}
\caption{
Dimensionless $D_{\rm T}^{(11)}(R,H)$ as a function 
of $R$ for various values of $H$. 
The red circles, blue squares and green triangles correspond 
to $H=0.1$, $1$ and $10$, respectively (color online).
The solid line corresponds to (\ref{eqn:DTSD}) representing 
the analytical expression for large $H$ limit.
The dashed line corresponds to the small $H$ limit.
}
\label{fig:DT1}
\end{figure}

For $H\gg1$, we can approximate $\coth(KH)\approx 1$ so that the 
$H$-dependence drops out.
In this limit, the integral can be analytically performed to yield 
\begin{eqnarray}
D_{\rm L}^{(11)}(R)= 
-\frac{2}{R^2} 
+\frac{\pi}{R}\left[ 
{\bf H}_1(R) - Y_1(R) 
\right],
\label{eqn:DLSD}
\end{eqnarray}
where ${\bf H}_n(t)$ are the Struve functions and $Y_n(t)$ are the 
Neumann functions.
Since the membranes are effectively isolated from each other
for large $H$, (\ref{eqn:DLSD}) coincides with the analytical 
expression obtained for a single membrane 
geometry~\cite{oppenheimer-09}.
For $R\ll1$, (\ref{eqn:DLSD}) has a logarithmic behavior, i.e.,
$D_{\rm L}^{(11)}(R\ll1)\approx \ln (2/R) -\gamma +0.5$ where 
$\gamma=0.5772\cdots$ is the Euler's constant.
In this regime, the outer solvent surrounding the membrane is 
unimportant and the membrane behaves effectively as a pure 2D fluid.
In the opposite $R\gg1$ limit, (\ref{eqn:DLSD}) decays 
algebraically $D_{\rm L}^{(11)}(R\gg1)\approx 2/R$.
On recasting in dimensional form, we see that this limiting 
expression is independent of the membrane viscosity $\eta$, 
and dependent only on the solvent viscosity $\eta_{\rm s}$.
When $H\ll 1$, on the other hand, (\ref{eqn:DL1}) becomes
$D_{\rm L}^{(11)}(R/2)/2$. 
The vanishing thickness of the solvent region ii results in a 
rescaling of $\nu^{-1}$ by a factor of two, and hence the resultant 
expression.
The solid and dashed lines in figure~\ref{fig:DL1} represent the 
above limiting cases of large and small $H$ limits, respectively.
It is thus observed that the presence of the second membrane also 
has a finite contribution to the coupling longitudinal diffusion 
coefficient within the same membrane.

Following the same argument, the transverse coupling diffusion 
coefficient between two particles within the same membrane 
is calculated according to 
\begin{eqnarray}
D_{\rm T}^{(11)}(R,H)
= 2  \int_0^\infty & {\rm d}K \,
\frac{1 + 2K +  \coth(KH)}{g(K,H)}
\nonumber\\ 
& \times \left[
J_0(KR) -\frac{J_1(KR)}{KR}
\right].
\label{eqn:DT1}
\end{eqnarray}
In figure~\ref{fig:DT1}, the variation of $D_{\rm T}^{(11)}(R,H)$ 
as a function of $R$ for various $H$ is plotted using 
numerical integration.
Similar to the longitudinal case, it is also a monotonically 
decreasing function of $R$. 
For $H\gg1$, (\ref{eqn:DT1}) can be analytically treated to have 
the form
\begin{eqnarray}
D_{\rm T}^{(11)}(R)
= &
\frac{2}{R^2} 
- \frac{\pi}{R} \left[ {\bf H}_1(R) -Y_1(R) \right] 
\nonumber \\
&+ \pi \left[{\bf H}_0(R) - Y_0(R)\right],
\label{eqn:DTSD}
\end{eqnarray}
which also coincides with the expression for the transverse 
coupling diffusion coefficient in a single 
membrane~\cite{oppenheimer-09}.
In this case, the asymptotic behaviors for small and large 
$R$ are 
$D_{\rm T}^{(11)}(R\ll1) \approx \ln(2/R) -\gamma -0.5$
and 
$D_{\rm T}^{(11)}(R\gg1)\approx 2/R^2$, respectively.
In the opposite limit of $H\ll1$, (\ref{eqn:DT1}) becomes 
$D_{\rm T}^{(11)}(R/2)/2$.
As before, the solid and dashed lines in figure~\ref{fig:DT1} 
show the limiting cases of large and small $H$, respectively.

\begin{figure}
\centering
\includegraphics[scale=0.45]{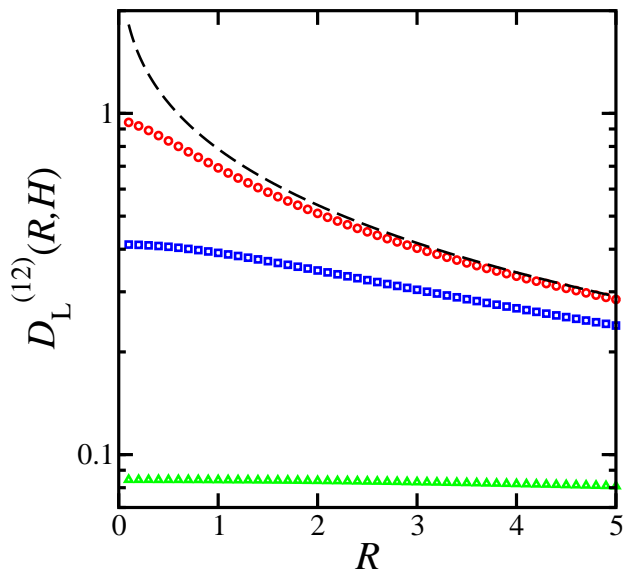}
\caption{
Dimensionless $D_{\rm L}^{(12)}(R,H)$ as a function 
of $R$ for various values of $H$. 
The red circles, blue squares and green triangles correspond 
to $H=0.1$, $1$ and $10$, respectively (color online).
The dashed line corresponds to the small $H$ limit.
}
\label{fig:DL2}
\end{figure}

Now we proceed to calculate the coupling diffusion coefficients
of two particles in different membranes.
In this case, the longitudinal coupling diffusion coefficient 
expressed in dimensionless units is 
\begin{equation}
D_{\rm L}^{(12)}(R,H)
= 2  \int_0^\infty {\rm d}K \,
\frac{{\rm cosech}(KH)}{g(K,H)} \,
\frac{J_1(KR)}{KR}.
\label{eqn:DL2}
\end{equation}
The functional dependence of $D_{\rm L}^{(12)}(R,H)$ on $R$ 
for various $H$ is shown in figure~\ref{fig:DL2}.
Since ${\rm cosech}(KH) \approx 0$ for $H\gg1$, the above 
integral vanishes for large inter-membrane distances
as seen by the green triangles ($H=10$) in figure~\ref{fig:DL2}.
This is a reasonable result as the membranes are effectively 
independent of each other.
In the limit of $H \ll 1$, (\ref{eqn:DL2}) results in 
$D_{\rm L}^{(11)}(R/2)/2$ which is plotted by the dashed line 
in figure~\ref{fig:DL2}.
As mentioned earlier, this function has an initial logarithmic
behavior followed by an asymptotic $1/R$-decay.

\begin{figure}
\centering
\includegraphics[scale=0.45]{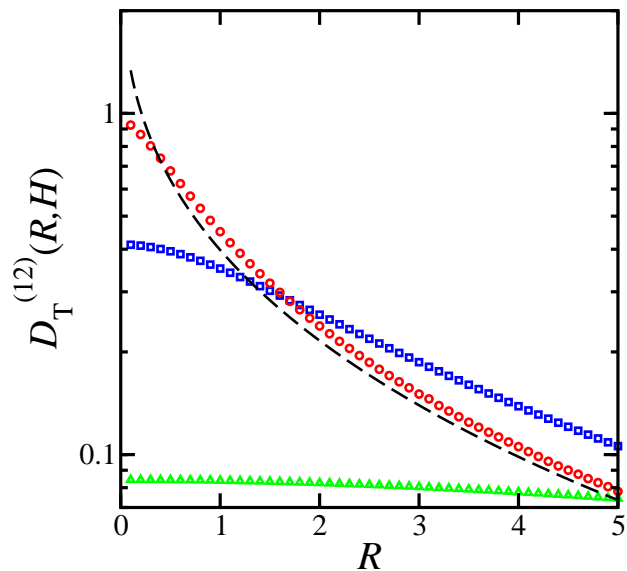}
\caption{
Dimensionless $D_{\rm T}^{(12)}(R,H)$ as a function 
of $R$ for various values of $H$. 
The red circles, blue squares and green triangles correspond 
to $H=0.1$, $1$ and $10$, respectively (color online).
The dashed line corresponds to the small $H$ limit.
}
\label{fig:DT2}
\end{figure}

The transverse coupling diffusion coefficient between two
particles in different membranes is given by
\begin{eqnarray}
D_{\rm T}^{(12)}(R,H)
= 2 \int_0^\infty & {\rm d}K \,
\frac{{\rm cosech}(KH)}{g(K,H)}
\nonumber \\
& \times \left[J_0(KR) -\frac{J_1(KR)}{KR}\right].
\label{eqn:DT2}
\end{eqnarray}
In figure~\ref{fig:DT2}, the variation of $D_{\rm T}^{(12)}(R,H)$ 
as a function of $R$ for various $H$ is plotted.
The above integral also vanishes when $H\gg1$ as expected 
for the decoupled membranes.
In the opposite limit of $H \ll 1$, (\ref{eqn:DT2}) results 
in $D_{\rm T}^{(11)}(R/2)/2$ as plotted by the dashed line in 
figure~\ref{fig:DT2}.

From figures~\ref{fig:DL2} and \ref{fig:DT2}, it can be observed 
that for large $H$, the coupling diffusion coefficients vanish.
This is due to the exponential decay of the cosech($KH$) term.
Up to $H=1$, the proximity to the second membrane leads to 
additional contributions to the coupling diffusion coefficients.
This is the main result of this work.
Using typical values for the solvent (water) and membrane 
(lipid bilayer) viscosities, the hydrodynamic screening length 
$\nu^{-1}$ can be estimated to be of the order of $\mu$m.
This implies that an adjacent membrane within this distance 
($H<1$) can have a strong bearing on the diffusion dynamics such as 
for typical stacked supported membrane experiments~\cite{kaizuka-04}.
Qualitatively, the presence of the second membrane can enhance the 
effective coupling diffusion.

Even though the model presented in this Communication captures the 
essential physics, it looks somewhat simplistic in several respects. 
We have neglected the finite size effect of the membrane inclusions 
which are known to modify the membrane response at small 
inter-particle distances~\cite{oppenheimer-09,oppenheimer-10}.
At distances much larger than the inclusion size, however,
these effects become unimportant.
Curvature effects of the membrane is significant when the radius of 
curvature is of the order of the hydrodynamic screening 
length $\nu^{-1}$~\cite{henle-08,henle-10}.
The extension of our work to membranes in spherical configuration 
is possible.
The out-of-plane fluctuations of the membrane lead to a reduction 
in the diffusion coefficient of proteins in the single 
membranes~\cite{naji-09,seifert-10}.
However, it is known that the presence of a substrate or the
second membrane would suppress the out-of-plane membrane 
fluctuations~\cite{sumithra-02,gov-04}.
Overall, we expect that fluctuations and the presence of a substrate 
will not qualitatively affect our results.

To summarize, we have calculated the longitudinal as well 
as transverse coupling diffusion coefficients of particles 
undergoing Brownian motion in a stacked double membrane geometry.
We obtained the additional contribution to the coupling diffusion
coming from the proximity to adjacent membranes.
As in biological systems such as tissues, the cells are rarely isolated, 
our results also imply that the proximity to neighboring cells can 
affect the diffusion of objects in cell membranes.


\ack
This work was supported by KAKENHI (Grant-in-Aid for Scientific
Research) on Priority Area ``Soft Matter Physics'' and Grant
No.\ 21540420 from the Ministry of Education, Culture, Sports, 
Science and Technology of Japan.

\section*{References}


\end{document}